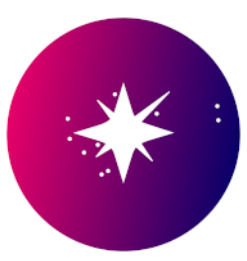

# Unveiling the Trans-Neptunian Region with the SKA

**Pablo Santos-Sanz**,[1] **Yücel Kilic**,[1] **Thomas G. Müller**,[2] **Emmanuel Lellouch**,[3] **Javier Moldón**[1] and **Jorma Harju**[4]

[1]*Instituto de Astrofísica de Andalucía-CSIC, Granada, Spain*
[2]*Max-Planck-Institut für extraterrestrische Physik (MPE), Garching, Germany*
[3]*LESIA - Observatoire de Paris, CNRS, France*
[4]*Department of Physics, University of Helsinki, Finland*

E-mail: psantos@iaa.es

We explore the potential of the SKA staged delivery AA4 (SKA-AA4) to detect and characterise the thermal emission of trans-Neptunian objects (TNOs) and Centaurs at centimetre wavelengths. These distant icy bodies preserve important information on the formation and evolution of the outer Solar System and provide a valuable link to planetary systems observed around other stars. The unprecedented sensitivity of SKA-AA4 will enable thermal detections of several of the brightest TNOs and Centaurs, extending radiometric studies beyond the capabilities of current radio facilities. Combined with shorter-wavelength observations and occultation measurements, these data will provide new constraints on spectral emissivity behaviour, thermophysical properties, and subsurface structure. At its highest angular resolutions, SKA-AA4 may partially resolve the largest systems, enabling investigations of surface heterogeneity, extended structures such as rings, and the partial resolution of some wide binary systems. We also assess the feasibility of radio occultation observations, which can provide independent constraints on object sizes, shapes, atmospheres, rings, or satellites. Together with observations from ALMA, JWST, and occultation campaigns, SKA-AA4 will establish centimetre-wavelength studies as a powerful new tool for investigating the thermal and structural properties of outer Solar System bodies.

## 1 Introduction

The trans-Neptunian region is populated by icy bodies considered among the most pristine remnants of the early stages of Solar System formation. These trans-Neptunian objects (TNOs), together with Centaurs and comets, are key to understanding the chemical composition, dynamical evolution, and collisional history of the outer Solar System. Their study also provides a direct link to the properties of protoplanetary discs observed around other stars, making them critical to broader questions of planetary system formation and evolution (see e.g., McClure et al., 2023; Marsset et al., 2023; Vernazza et al., 2025; Pinilla-Alonso et al., 2025; Xie et al., 2025).

Over the past two decades, thermal observations from space-based missions such as Spitzer (Stansberry et al., 2008), Herschel (Müller et al., 2009; Lellouch et al., 2013), WISE (Bauer et al., 2013), and, more recently, JWST (Kiss et al., 2024; Pinilla-Alonso et al., 2025; Licandro et al., 2025; De Prá et al., 2025; Brunetto et al., 2025), together with ground-based facilities such as ALMA (Moullet et al., 2011; Gerdes et al., 2017; Brown and Butler, 2017, 2018; Lellouch et al., 2017), have provided crucial constraints on the physical properties and thermal behaviour of TNOs and Centaurs. A comprehensive overview of these multi-instrument thermal studies and their results is presented by Müller et al. (2020), which collectively form the basis for our current understanding of the thermal emission of these distant bodies. These observatories have yielded essential insights into the physical nature of several large bodies; however, they are generally limited to shorter wavelengths (far-infrared and submillimetre), which probe only the very surface layers of these cold, distant objects. Their wavelength coverage does not extend into the centimetre domain, where deeper subsurface regions can be investigated. Moreover, with no forthcoming space-based facility capable of accessing the thermal peak of these objects (at ~50–100 $\mu$m), further progress will depend on ground-based observations at longer wavelengths. Extending thermal studies into the centimetric regime will therefore allow access to subsurface layers, potentially revealing structural and compositional stratification that remains undetectable at higher frequencies.

The Square Kilometre Array (SKA), with its unprecedented sensitivity in the radio domain, will enable a new observational window into the outer Solar System. In particular, the SKA-AA4 will offer the sensitivity required to detect the faint thermal emission of several of the brightest TNOs and Centaurs, especially at its shortest operational wavelengths. These detections will provide important constraints on spectral emissivity behaviour and subsurface thermal properties of TNOs and Centaurs, particularly when interpreted through thermophysical models and combined with shorter-wavelength observations and occultation measurements.

In addition to thermal studies, SKA observations may also enable radio-occultation measurements of compact background sources by TNOs and Centaurs. Although relatively rare, such events can provide precise constraints on object sizes, shapes, rings, atmospheres, or satellites, complementing thermal radiometry with independent geometric information.

Ultimately, SKA observations will provide new constraints on the formation and evolutionary pathways of outer Solar System populations, while also contributing to studies of planetary migration scenarios and the dynamical evolution of the trans-Neptunian region (DeMeo and Carry, 2014).

## 2 Methodology

To assess the feasibility of detecting the thermal emission of TNOs and Centaurs with the SKA-AA4, we combined thermal modelling, based on simplified radiometric approaches, with previous observational thermal data. Our approach focuses on predicting flux densities at centimetric wavelengths for a representative sample of the 23 brightest known TNOs and Centaurs, based on current knowledge of their sizes, albedos, and heliocentric distances.

Thermal fluxes were computed using simple thermal models, specifically the Standard Thermal Model (STM) in the framework of the Near-Earth Asteroid Thermal Model (NEATM; Harris 1998; Lebofsky et al. 1986). This approach assumes instantaneous thermal equilibrium between insolation and re-radiation, with a beaming parameter adjusted to reproduce the observed fluxes. Input parameters include the object's effective diameter, geometric albedo, phase integral (Verbiscer et al., 2022), and an assumed bolometric emissivity. The surface is treated as a homogeneous, airless regolith with low thermal conductivity, representative of icy, porous material typical of TNOs and Centaurs. Unlike thermophysical models, the STM does not explicitly solve the heat conduction equation but provides a robust first-order estimate of thermal emission for distant Solar System bodies, given the limited constraints currently available on their surface properties.

The adopted values of the beaming parameter ($\eta = 1.2 \pm 0.35$) and phase integral ($q = 0.39$) are intended to provide representative first-order estimates for the sample as a whole, rather than object-specific thermophysical solutions. Although phase integrals and thermal parameters are known for some individual TNOs and Centaurs (e.g. Verbiscer et al., 2022), homogeneous values were adopted here in order to maintain a consistent exploratory framework for detectability estimates across the full sample. We note that uncertainties associated with emissivity behaviour at centimetre wavelengths are likely to dominate over moderate variations in these parameters.

For this study, a spectral emissivity of 0.7 was adopted at radio wavelengths, consistent with values derived from previous fits to thermal data obtained with Herschel (PACS and SPIRE), Spitzer (MIPS), and ALMA (Fornasier et al., 2013; Lellouch et al., 2017; Brown and Butler, 2017). This assumption provides a reasonable first-order approximation for extrapolating thermal models into the centimetre regime. For high-albedo objects, the effective absorbed energy may differ from simplified visible-band approximations due to the spectral dependence of surface reflectance, particularly at red and near-infrared wavelengths. These effects are not explicitly included in the simplified STM approach adopted here.

Nevertheless, it is not yet established that an emissivity of 0.7 remains valid at such long wavelengths. Laboratory and observational studies suggest that emissivity can decrease further at centimetre wavelengths due to enhanced volume scattering or dielectric effects in porous icy surfaces (Webster and Johnston, 1989; Müller and Barnes, 2007). The resulting model fluxes were therefore extrapolated to the AA4 wavelength coverage with caution, focusing on the shortest operational wavelengths (2.3 cm, 2.8 cm, and 4.6 cm) within Bands 5b and 5a, where this assumption is most likely to remain applicable. Longer-wavelength data points were excluded because of the expected rapid drop in flux density and the increasing contribution from the sky background.

The predicted fluxes were compared with the expected continuum sensitivity of SKA-AA4 in each wavelength, assuming a 5-hour integration time and a 5$\sigma$ detection threshold. These detection threshold estimates are based on the SKA sensitivity calculator for the staged delivery AA4[1] (see Table 1). This approach allows us to define the region of parameter space in which SKA observations, particularly when combined with shorter-wavelength radiometry and occultation measurements, can provide meaningful constraints on emissivity behaviour and thermal inertia, while improving constraints on size and albedo.

**Table 1:** Estimated SKA-AA4 sensitivities at 5$\sigma$ for 1 h and 5 h of integration derived from the SKA sensitivity calculator[1]. $\lambda_c$ and $\lambda_{range}$ denote the central wavelength and wavelength range, respectively, while $\nu_c$ and $\nu_{range}$ indicate the corresponding central frequency and frequency range.

| $\lambda_c$ [cm] | $\lambda_{range}$ [cm] | $\nu_c$ [GHz] | $\nu_{range}$ [GHz] | **SKA-AA4 (1h)** [µJy] | **SKA-AA4 (5h)** [µJy] |
|---|---|---|---|---|---|
| 4.6 | 3.5–6.5 | 6.55 | 4.60–8.50 | 2.72 | 1.22 |
| 2.8 | 2.3–3.6 | 10.80 | 8.30–13.30 | 3.12 | 1.39 |
| 2.3 | 1.9–2.9 | 12.90 | 10.40–15.40 | 3.54 | 1.58 |

Target selection was based on apparent visible brightness and estimated size, which, when combined with known distances and assumed surface properties, provide a reasonable proxy for the expected thermal flux at radio wavelengths. Previous experience from thermal radiometry programmes such as *TNOs are Cool* (e.g. Müller et al., 2020) has shown that the optically brightest known TNOs generally correspond to the largest and most thermally detectable objects, despite some expected exceptions related to albedo and emissivity variations. In practice, visible brightness serves as a useful first-order proxy for identifying large and potentially detectable targets at centimetre wavelengths.

The final list includes a mix of dynamical classes—classical TNOs, plutinos, scattered disc objects, and Centaurs—with diameters typically larger than ~200 km, as well as several smaller but nearby or unusually bright objects. The selected targets were prioritised not only according to their expected detectability but also based on the availability of previous thermal measurements from Spitzer, Herschel, and ALMA, together with stellar occultation results for many of the objects, including several multi-chord events providing accurate size, 3D shape, and ring constraints. Figure 1 presents representative thermal models for some of these targets, based on previous infrared and submillimetre observations and extrapolated to centimetric wavelengths. These examples illustrate the type of outputs obtained from our modelling approach.

## 3 Scientific Objectives

The trans-Neptunian region offers a unique opportunity to investigate the physical properties and evolutionary history of primitive Solar System bodies. The primary objective of this study is to assess the scientific potential of the SKA-AA4 to characterise trans-Neptunian objects (TNOs) and Centaurs through their thermal emission at centimetre wavelengths.

[1] https://sensitivity-calculator.skao.int/

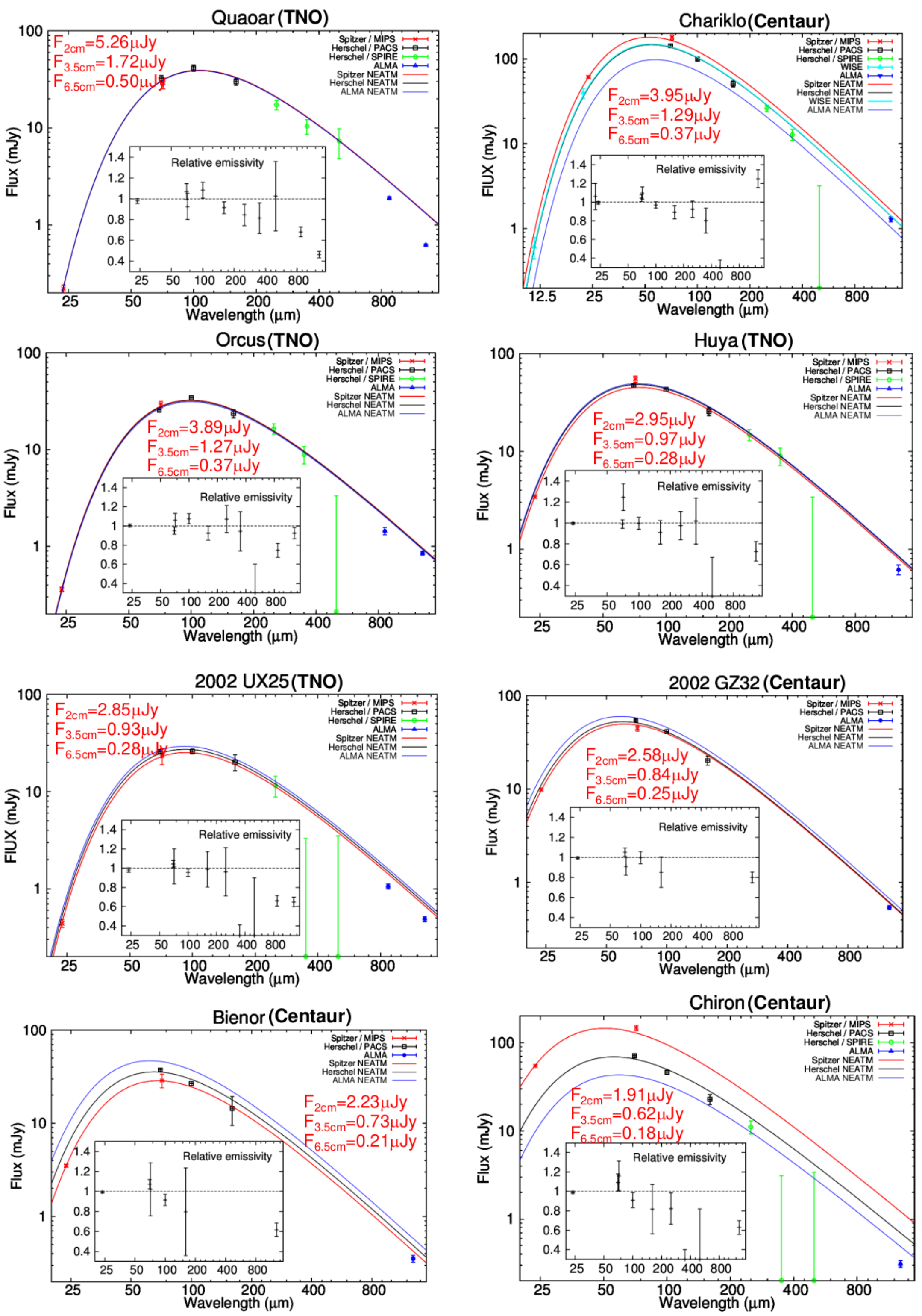

**Figure 1:** Thermal models for a selection of large TNOs and Centaurs, based on Spitzer, Herschel, and ALMA data. Insets show the spectral emissivity for each object. Expected SKA fluxes at 2, 3.5, and 6.5 cm (in red) are extrapolated assuming a relative spectral emissivity of $\epsilon$ = 0.7. Adapted from Lellouch et al. (2017) and subsequent works.

Thermal observations at these long (centimetric) wavelengths are particularly valuable for probing subsurface layers, enabling the detection of thermal emission originating from depths inaccessible to far-infrared or submillimetre instruments. The centimetre regime is sensitive to thermal radiation emerging from depths of a few metres, corresponding to the electromagnetic skin depth at these frequencies (Ries and Janssen, 2015). Measuring spectral emissivity in this wavelength range will therefore allow us to explore physical parameters, such as layering, porosity, dielectric constant, and composition, beneath the surface. These properties are directly linked to the effects of irradiation, stratification, and sublimation processes, as highlighted in the *DiSCo* programme (e.g. Pinilla-Alonso et al., 2021, 2025), and are essential for understanding the long-term evolution and surface renewal of these distant icy worlds. Variations in emissivity may also reveal differences in regolith structure, composition, or the efficiency of volume scattering, as observed on icy satellites in the outer Solar System (Black et al., 2007; Ostro et al., 2006; Ries and Janssen, 2015). In turn, such measurements, particularly when combined with shorter-wavelength radiometry and occultation constraints, will improve thermophysical models and provide more robust estimates of effective diameters, albedos, thermal inertia, and emissivity behaviour.

Moreover, accurate determinations of size and albedo, when combined with mass estimates from binary systems, will allow the derivation of bulk densities for an increasing number of TNOs (e.g. Mommert et al., 2012; Santos-Sanz et al., 2012; Vilenius et al., 2012, 2014; Fornasier et al., 2013). This will help refine the population's size distribution and improve our understanding of its formation and dynamical evolution.

Another potential objective is the detection and characterisation of ring systems around TNOs and Centaurs. The discovery of dense icy rings around Haumea (Ortiz et al., 2017), Quaoar (Morgado et al., 2023; Pereira et al., 2023), Chariklo (Braga-Ribas et al., 2014), and Chiron (Ortiz et al., 2015; Braga-Ribas et al., 2023; Ortiz et al., 2023; Pereira et al., 2025), suggests that ring formation in the outer Solar System may be more common than previously thought. If these rings are dominated by sub-centimetre to metre-sized icy particles, they could, in principle, produce detectable thermal emission with SKA-AA4. However, a detailed assessment is still required to determine whether the SKA's sensitivity and spatial resolution will be sufficient to detect or resolve such structures, given the trade-off between higher sensitivity and lower angular resolution. The detection of dense icy rings would provide valuable constraints on particle size distributions, albedo, and dynamical stability, whereas tenuous or dusty rings, such as those of the giant planets, are expected to remain below the centimetric sensitivity limits of current facilities.

A secondary goal is to evaluate the feasibility of detecting radio occultations of compact background sources by TNOs and Centaurs. Although such events are rare and observationally challenging, they offer high-precision constraints on object size and shape. Occultations can also reveal the presence of tenuous atmospheres, ring systems, or binary companions. The combination of occultation-derived geometry with thermal flux measurements has the potential to significantly enhance our understanding of the bulk and surface properties of these remote worlds.

In addition, SKA-AA4 observations may provide the first opportunity to obtain rotational lightcurves for selected large TNOs at centimetre wavelengths, thereby probing surface or subsurface heterogeneities. Such measurements would complement optical and infrared lightcurves by constraining variations in emissivity or dielectric properties across the surface, potentially revealing local compositional or structural differences.

An intriguing prospect is the detection of thermally active TNOs. Recent JWST observations suggest that activity might be ongoing on the dwarf planet Makemake (Kiss et al., 2024), and similar phenomena could, in principle, be investigated with the SKA if the emission contrast is sufficiently high. Although micron-sized particles in cryovolcanic plumes would remain invisible in the centimetric range, SKA observations could still detect temperature anomalies caused by exposed or recently resurfaced areas where coarser grains have formed or where the dielectric properties differ from the surrounding terrain. Comparable anomalies have been reported on icy satellites such as Enceladus: Very Large Array (VLA) measurements at 13 cm show it as a faint thermal emitter consistent with cold water ice (Black et al., 2007), while Cassini-RADAR observations at 2.17 cm revealed a thermal anomaly on its leading hemisphere attributed to a younger surface (Ostro et al., 2006; Ries and Janssen, 2015). These results demonstrate the diagnostic power of centimetre-wavelength studies for distinguishing surface age, texture, and composition.

While JWST is expected to provide valuable data in the near- and mid-infrared up to 28 $\mu$m (Rieke et al., 2015), it is not optimal for radiometric measurements of TNOs or Centaurs. ALMA and ngVLA can probe into the millimetre domain, but even their capabilities remain limited to the brightest and largest objects. Only SKA-AA4, with its exceptional sensitivity at centimetre wavelengths, will be able to detect thermal emission from a significantly larger sample of mid-sized TNOs and Centaurs. Such an expanded sample will enable broader comparative studies, improving our capacity to investigate their thermal behaviour, internal structure, and evolutionary pathways.

## 4 Results

In this section, we present the results of our investigation into the capabilities of the SKA-AA4 to characterise TNOs and Centaurs through centimetre-wavelength observations. The results are structured into two complementary studies, each focusing on a distinct observational approach.

The first part (§4.1) addresses the detectability of the thermal emission from TNOs and Centaurs. We use thermal modelling to estimate flux densities at selected SKA bands for a sample of the 23 brightest known objects. These predictions are compared to the expected sensitivity of SKA-AA4 to determine which targets are likely to be detectable under realistic observing conditions. This analysis also demonstrates the potential of SKA observations not only to constrain physical parameters, such as spectral emissivity and thermal inertia, but also to probe subsurface structure in a way inaccessible to shorter-wavelength facilities.

The second part (§4.2) explores the feasibility of detecting radio occultations of compact background sources by TNOs. Although such events are inherently rare, the exceptional sensitivity and spatial coverage of the SKA could enable their detection under favourable circumstances. We examine the diffraction signatures expected during these occultations, discuss the observing requirements, and outline the scientific returns in terms of object size and shape, as well as the potential detection of rings, atmospheres, or satellites.

### 4.1 Thermal Emission Detectability

Observations of the thermal emission from trans-Neptunian objects (TNOs) and Centaurs at centimetric wavelengths provide important diagnostics of their thermophysical properties, spectral emissivity behaviour, and surface thermal regimes. These measurements are compelling when interpreted through thermal (TMs) and thermophysical models (TPMs), as demonstrated in previous studies using Spitzer, Herschel, and ALMA data (Stansberry et al., 2008; Lellouch et al., 2013; Santos-Sanz et al., 2017; Müller et al., 2020; Lellouch et al., 2017; Brown and Butler, 2017). With SKA-AA4, we aim to extend this approach to longer wavelengths, probing subsurface layers that remain inaccessible to far-infrared or submillimetre instruments.

Our simulations show that while SKA-AA* would only be capable of detecting the thermal emission from a few of the largest TNOs, the increased sensitivity of the SKA-AA4 enables detection of a significantly broader population, including several of the largest TNOs and Centaurs. In particular, we consider a sample composed of the 23 brightest known objects in the trans-Neptunian region.

Using a 5-hour integration time and a $5\sigma$ detection threshold, we find that a substantial fraction of these targets are within the reach of SKA-AA4 at its shortest operating central wavelength (2.3 cm), corresponding to the upper limit in frequency of Band 5b. At this wavelength, SKA-AA4 is expected to detect a sizeable subset of the brightest TNOs and Centaurs above the 1.58 $\mu$Jy threshold, making it the most productive band for thermal detections. The adopted spectral emissivity value of $\epsilon = 0.7$, consistent with previous ALMA-based studies (Lellouch et al., 2017; Brown and Butler, 2017), allows us to extrapolate from existing thermal data and estimate the expected SKA fluxes with reasonable confidence. Representative thermal models for some of the brightest TNOs and Centaurs, along with their predicted centimetric fluxes, are illustrated in Figure 1.

The full results for the sample are presented in Table 2, which illustrates the diversity of thermal emission levels accessible to SKA-AA4 under realistic observing conditions. Across all wavelengths considered (2.3, 2.8, and 4.6 cm), 14 of the 23 objects (61%) are detectable at $5\sigma$ in 5 h integrations. At 2.8 cm, 12 targets (52%) remain above the corresponding sensitivity threshold, while at 4.6 cm, only Pluto and Haumea (9% of the sample) are expected to be detected. These statistics confirm the strong wavelength dependence of detectability for TNOs and Centaurs, with shorter centimetre wavelengths providing the most favourable conditions for robust detections and meaningful thermophysical and emissivity constraints (see Table 3 for a detailed breakdown by band).

Thermal observations at these wavelengths open the door to new investigations of spectral emissivity behaviour in the centimetre regime. It is well known that the emissivity of TNOs tends to decrease at long wavelengths, often in a wavelength-dependent fashion. Several mechanisms have been proposed to explain this effect, including subsurface sounding of colder layers (Lellouch et al., 2017), dielectric reflection of thermal radiation (Fornasier et al., 2013), and volume scattering by voids or inhomogeneities with size scales on the order of $\lambda/4\pi$ (Brown and Butler, 2017).

By obtaining multi-wavelength continuum measurements with SKA-AA4, particularly when combined with existing Spitzer, Herschel, and ALMA data, we can place meaningful constraints on these processes and thus gain insight into the physical structure and composition of the upper layers of these distant bodies. For objects with well-characterised sizes and thermal inertias from prior studies, SKA observations will serve as a benchmark for emissivity behaviour. In turn, this may help extend thermophysical studies to a broader population of fainter TNOs and Centaurs for which little or no thermal information currently exists.

**Table 2:** 23 largest TNOs: name, type, diameters, heliocentric distance ($r_h$) as of October 2025 (MPC), and flux densities extrapolated from thermal models at 2.3–4.6 cm (µJy), together with brightness temperatures. The thermal fluxes were computed using the Standard Thermal Model (STM), assuming a phase integral of $q = 0.39$ (e.g. Lebofsky et al., 1986; Harris, 1998) and a beaming parameter of $\eta = 1.2 \pm 0.35$, consistent with values derived for TNOs from Lellouch et al. (2013). Green cells indicate expected $\geq 5\sigma$ detections with SKA-AA4 after 5 h of integration, based on the sensitivity estimates for 2.3, 2.8, and 4.6 cm (see Table 1).

| Object | Type | D [km/mas] | $r_h$ [au] | $F_{2.3cm}$ [µJy] | $F_{2.8cm}$ [µJy] | $F_{4.6cm}$ [µJy] | $T_b$ [K] |
|---|---|---|---|---|---|---|---|
| Pluto | TNO Plut. | 2380/100 | 35.38 | 35.7 | 25.1 | 9.25 | 36.5 |
| Eris | TNO Det. | 2326/33 | 95.54 | 2.25 | 1.58 | 0.58 | 22.1 |
| Haumea | TNO Clas. | 1595/43 | 49.84 | 4.94 | 3.48 | 1.29 | 28.3 |
| Gonggong | TNO SDO | 1535/24 | 89.67 | 1.44 | 1.02 | 0.38 | 26.2 |
| Makemake | TNO Clas. | 1430/38 | 52.72 | 3.51 | 2.46 | 0.91 | 26.6 |
| Quaoar | TNO Clas. | 1071/34 | 42.63 | 3.88 | 2.73 | 1.01 | 36.4 |
| Orcus | TNO Plut. | 960/28 | 48.00 | 2.87 | 2.01 | 0.74 | 40.8 |
| Mani | TNO Clas. | 934/27 | 46.05 | 2.68 | 1.89 | 0.70 | 38.3 |
| Salacia | TNO Clas. | 909/28 | 45.32 | 2.86 | 2.01 | 0.73 | 37.2 |
| Sedna | TNO Det. | 906/14 | 83.07 | 0.58 | 0.40 | 0.15 | 30.0 |
| Uni | TNO Clas. | 695/24 | 39.57 | 2.10 | 1.48 | 0.55 | 40.0 |
| Ritona | TNO Clas. | 679/23 | 40.58 | 2.06 | 1.45 | 0.54 | 42.9 |
| Varuna | TNO Clas. | 668/21 | 44.23 | 1.41 | 1.00 | 0.38 | 34.9 |
| 2003 $VS_2$ | TNO Plut. | 523/19 | 36.94 | 1.33 | 0.93 | 0.34 | 36.8 |
| Huya | TNO Plut. | 458/22 | 29.27 | 2.18 | 1.53 | 0.56 | 47.7 |
| Lempo | TNO Plut. | 393/17 | 31.08 | 1.44 | 1.01 | 0.38 | 50.5 |
| 2002 $TX_{300}$ | TNO Clas. | 286/9 | 43.43 | 0.27 | 0.18 | 0.06 | 32.1 |
| Chariklo | Centaur | 241/21 | 17.59 | 2.91 | 2.04 | 0.75 | 68.4 |
| 2002 $GZ_{32}$ | Centaur | 237/17 | 18.50 | 1.90 | 1.34 | 0.50 | 66.4 |
| Chiron | Centaur | 210/15 | 18.44 | 1.40 | 0.99 | 0.36 | 66.3 |
| Bienor | Centaur | 199/16 | 13.30 | 1.64 | 1.15 | 0.42 | 63.5 |
| Typhon | TNO SDO | 185/13 | 26.31 | 1.06 | 0.74 | 0.27 | 59.2 |
| Okyrhoe | Centaur | 35/7 | 9.27 | 0.52 | 0.37 | 0.14 | 119.0 |

Beyond the main results presented here, SKA observations may also provide insights into additional aspects of TNO and Centaur systems, including multiple systems, surface heterogeneities, and ring material. Dense icy rings have already been identified around several of these objects (Ortiz et al., 2017; Morgado et al., 2023; Pereira et al., 2023; Braga-Ribas et al., 2014; Ortiz et al., 2015; Pereira et al., 2025), and future centimetre-wavelength observations may help further investigate the structural diversity and complexity of these environments under favourable observing conditions.

**Table 3:** Number of detectable objects at $5\sigma$ in 5 h integrations with SKA-AA4 at three wavelengths. Percentages are relative to the total sample of 23 objects.

| $\lambda$ [cm] | Threshold (5 h) [$\mu$Jy] | N obj. (%) | Detected objects |
|---|---|---|---|
| 2.3 | 1.58 | 14 (61%) | Pluto, Eris, Haumea, Makemake, Quaoar, Orcus, Mani, Salacia, Uni, Ritona, Huya, Chariklo, 2002 $GZ_{32}$, Bienor |
| 2.8 | 1.39 | 12 (52%) | Pluto, Eris, Haumea, Makemake, Quaoar, Orcus, Mani, Salacia, Uni, Ritona, Huya, Chariklo |
| 4.6 | 1.22 | 2 (9%) | Pluto, Haumea |

### 4.2 Potential for Radio Occultation Studies

Radio occultation observations represent a powerful, though technically demanding, method to investigate TNOs and Centaurs. These events, which occur when a TNO passes in front of a compact background radio source, can yield high-precision information on the size, shape, position, binary nature, and potentially the presence of atmospheres or ring systems associated with the occulting body (Lehtinen et al., 2016; Harju et al., 2018). Although the SKA was not explicitly designed for this purpose, its exceptional sensitivity and resolution make it a promising facility for detecting such events, provided the occultation geometry is favourable.

Most occultations by TNOs will occur in the Fraunhofer diffraction regime, where the extent of the diffraction fringes greatly exceeds the object's geometrical silhouette (Trahan and Hyland, 2014). This spreading of the diffraction pattern increases the effective detection cross-section, though it simultaneously reduces the signal strength. The key quantity governing the detectability of these events is the Fresnel number, defined as $F = a^2/(z\lambda)$, where $a$ is the radius of the object, $z$ its distance from the observer, and $\lambda$ the observing wavelength.

At a frequency of 15 GHz ($\lambda = 2$ cm), relevant for SKA-AA4, the Fresnel number for a 300 km TNO at 50 au is $F = 0.15$. This corresponds to a regime where both amplitude and phase variations due to diffraction are observable. Figure 2 illustrates the intensity and phase structure of such an occultation, including the time series of the visibility as the shadow traverses the array. A clear Arago spot appears at the moment of closest approach, surrounded by diffraction fringes that encode the size and structure of the occulting object.

The detectability threshold for such events can be estimated using SKA performance figures. Assuming an SKA-AA4 effective area-to-system temperature ratio of $A_e/T_{sys} \approx 450\,\mathrm{m^2\,K^{-1}}$, and a correlator integration time of 0.1 s, a TNO of diameter $D \sim 300$ km at 50 au can be detected with a signal-to-noise ratio greater than 50 at $\lambda = 2$ cm. In general, the minimum detectable diameter at this wavelength scales as $a_{\min} \sim 20\sqrt{z\,[\mathrm{au}]}$ km, if we request a signal-to-noise ratio of 50 in the amplitude variation, implying that many of the largest and brightest TNOs fall within reach of occultation studies.

Technical challenges remain significant. Because such events are both rare and challenging to predict, they demand highly accurate ephemerides and careful coordination with other facilities. The scarcity of suitable K-band calibrators also limits the number of viable background sources. Phase calibration is critical: the ideal strategy would involve using Very Long Baseline Interferometry (VLBI), with part of the array observing a phase reference outside the occultation path. If VLBI is unavailable, alternative calibration can be achieved by tracking instrumental and atmospheric phase drifts before and after the event, since their timescales are typically longer than the occultation duration.

Despite these challenges, radio occultations provide a valuable complement to thermal observations. When combined with radiometric fluxes from SKA or other facilities, occultation-derived sizes enable accurate determinations of albedos and provide improved constraints on thermal inertias through thermal modelling. Furthermore, diffraction patterns can reveal departures from spherical symmetry, binary companions, or ring systems (Braga-Ribas et al., 2014; Ortiz et al., 2015, 2017; Morgado et al., 2023), or constrain tenuous atmospheres, as previously suggested for several Centaurs and TNOs (Ortiz et al., 2012), and more recently strengthened by the detection of atmospheric signatures around a mid-sized TNO beyond Pluto (Arimatsu et al., 2026). The SKA thus opens a unique window for exploring the physical and dynamical properties of outer Solar System bodies through high-precision radio occultation science.

## 5 Discussion

The advent of the SKA-AA4 represents a significant milestone for Solar System science, offering unprecedented sensitivity at centimetre wavelengths. By extending thermal measurements beyond the far-infrared and submillimetre regimes previously explored with facilities such as ALMA, Herschel, and Spitzer, SKA-AA4 will probe subsurface layers and dielectric properties that remain largely unexplored.

Our results show that SKA-AA4 will enable thermal detections of a significantly broader sample of TNOs and Centaurs than currently accessible at radio wavelengths, including several mid-sized objects. Multi-wavelength observations in the centimetre regime will provide important constraints on the behaviour of spectral emissivity and subsurface thermal properties. However, because centimetre emissivities remain poorly constrained for most TNOs, SKA observations alone will generally not provide unique solutions for diameter, albedo, and thermal inertia. Instead, their full diagnostic power will emerge when combined with shorter-wavelength radiometry and occultation-derived size measurements.

The angular resolution achievable with SKA-AA4 will also play a key role in assessing the spatially resolved emission from the brightest TNOs and Centaurs. Depending on the frequency, the synthesised beam can range from ~0.05” to 0.02” (50–20 mas) over the 4.6–2.3 cm wavelength range, when the array is weighted for resolution. Under such conditions, Pluto (∼100 mas) and the Haumea ring system (∼120 mas) could become at least partially resolved, while other large TNOs such as Eris, Makemake, or Quaoar (30–40 mas) would likely remain unresolved. Chariklo’s ring system, with a total extent of ~70 mas, lies close to the instrumental resolution limit. However, achieving such angular resolutions comes at the expense of sensitivity, and dedicated simulations will be required to assess whether sufficient signal-to-noise ratios can be maintained to reliably image these structures.

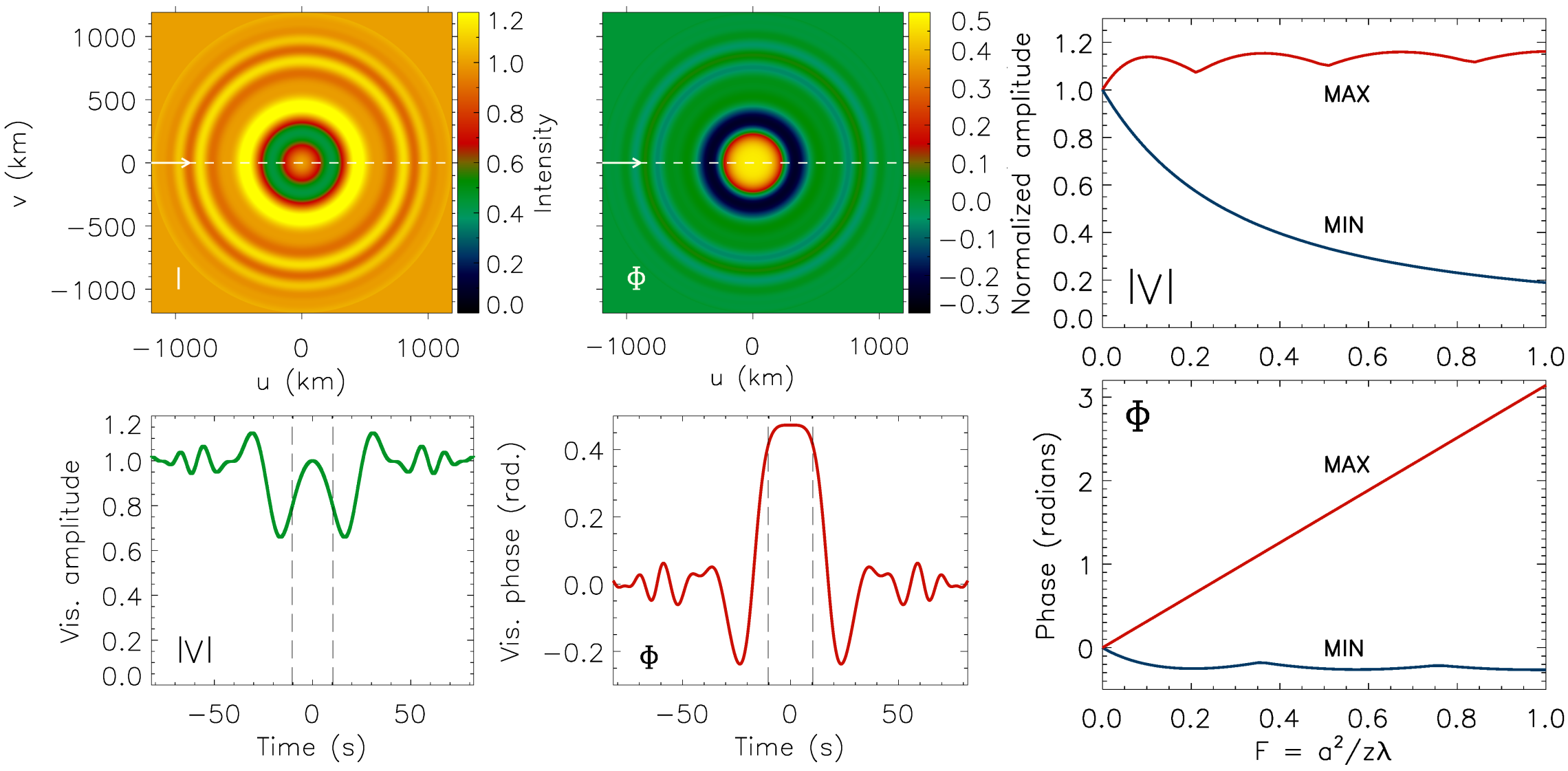

**Figure 2: Left:** Simulated diffraction patterns observed on Earth during a central radio occultation by a trans-Neptunian object (TNO) with a circular geometry. The TNO is assumed to have a diameter of 300 km and to be located at a heliocentric distance of 50 au. Observations are performed at 15 GHz ($\lambda = 2$ cm), corresponding to a Fresnel number of $F = 0.15$. The top panels display the two-dimensional intensity and phase patterns projected onto a plane perpendicular to the line of sight. The bottom panels show the corresponding time series of visibility amplitude and phase as recorded by a radio interferometer, as the diffraction pattern (the "shadow") moves across the instrument. Here, the occultation is central as indicated by the dashed lines in the top panels. The vertical dashed lines in the lower plots mark the expected immersion and emersion times at optical wavelengths. The central brightness peak corresponds to the Arago spot, associated with the point of closest approach. **Right:** Maximum and minimum values of the visibility amplitude and phase as a function of the Fresnel number.

These capabilities enable investigations of surface or subsurface heterogeneity through local variations in brightness temperature, emissivity, or dielectric properties. In addition, centimetre-wavelength observations may prove particularly sensitive to water-ice-rich terrains owing to the low absorption and strong volume scattering properties of water ice at these wavelengths. This capability may be especially relevant for members of the Haumea family (Vilenius et al., 2018), which are characterised by unusually high albedos and water-ice-rich surfaces, making them particularly attractive targets for centimetre-wavelength emissivity studies.

SKA-AA4 observations may also provide new opportunities for studying binary and multiple TNO systems. For the widest and brightest binaries, partial spatial separation of the components could become feasible, while multi-epoch observations may constrain flux ratios and approximate size ratios between components. Combined with orbital solutions, such measurements would improve density estimates and provide valuable constraints on binary formation and collisional evolution scenarios.

Radio occultations of compact background sources by TNOs and Centaurs represent another promising application of SKA. Although technically challenging and relatively rare, such events can provide highly accurate constraints on object sizes, shapes, rings, atmospheres, or binary companions. The high time and phase resolution of the SKA will enable detection of diffraction signatures during favourable occultation events, thereby complementing thermal radiometry with independent geometric information. Combined with centimetre-wavelength radiometry, these occultation measurements will also provide valuable constraints on the collisional and dynamical evolution of the outer Solar System.

Ring systems around Centaurs and TNOs may also provide valuable information on particle size distributions and thermal properties at centimetre wavelengths. These considerations primarily apply to rings dominated by macroscopic icy particles, which are expected to dominate the centimetric thermal emission. However, transient populations of micron- or submicron-sized dust grains may also exist in some systems, as suggested by recent observations of evolving material around the Centaurs Chiron and Chariklo (Pereira et al., 2025; Santos-Sanz et al., 2025). Although such fine-grained dust is expected to contribute only weakly at centimetre wavelengths, transient dusty activity could still generate detectable local thermal or emissivity anomalies under favourable conditions.

SKA-AA4 observations will strongly benefit from synergies with JWST, ALMA, and coordinated stellar occultation campaigns. In addition, the Legacy Survey of Space and Time (LSST) at the Vera C. Rubin Observatory is expected to dramatically expand the known trans-Neptunian population over the coming decade. Current projections suggest that LSST may discover on the order of 30 000–40 000 new TNOs (e.g. Ivezić et al., 2019; LSST Solar System Science Collaboration, 2019), compared to the ~6 000 objects currently known. Although most newly discovered objects will remain too faint for centimetre thermal detections with SKA-AA4, the greatly expanded census will substantially increase the number of potential radio occultation targets accessible to the SKA, enhancing opportunities for high-precision determinations of size, shape, rings, and possible atmospheres. The joint analysis of data from SKA, JWST, ALMA, and occultation campaigns will also help place the architecture and evolutionary pathways of the outer Solar System in the broader context of planetary systems observed around other stars.

## 6 Conclusions

The SKA-AA4 will open a new observational window for the study of TNOs and Centaurs at centimetre wavelengths. Our simulations indicate that SKA-AA4 will enable thermal detections of several of the largest and brightest outer Solar System bodies, extending thermal studies into a wavelength regime that remains largely unexplored for these objects.

Combined with complementary observations from facilities such as JWST, ALMA, Herschel, and Spitzer, as well as stellar and radio occultation measurements, SKA-AA4 observations will provide important constraints on spectral emissivity behaviour, subsurface thermal properties, and the physical structure of distant icy bodies. In particular, the centimetre regime will offer unique opportunities to investigate surface and subsurface properties, ring systems, binary systems, and other forms of structural complexity in the outer Solar System.

Future modelling and simulated observations will nevertheless remain essential for optimising observing strategies and assessing the practical detectability of partially resolved systems and small-scale features. Overall, SKA-AA4 will establish centimetre-wavelength observations as a powerful new tool for exploring the formation, evolution, and diversity of outer Solar System bodies.

## Acknowledgements

P.S.-S., Y.K. and J.M. acknowledge financial support from the Severo Ochoa grant CEX2021-001131-S funded by MCIN/AEI/10.13039/501100011033. P.S.-S. and Y.K. acknowledge financial support from the Spanish I+D+i project PID2022-139555NB-I00 (TNO-JWST) funded by MCIN/AEI/10.13039/501100011033.